# Distance-Dependent Evolution of Electronic States in Kagome-Honeycomb Lateral Heterostructures in FeSn

Tuan Anh Pham,[1,#] Seoung-Hun Kang,[2,#] Yasemin Ozbek,[1] Mina Yoon,[2,*] and Pengpeng Zhang[1,*]

[1]Department of Physics and Astronomy, Michigan State University, East Lansing, MI, 48824, U.S.A.

[2]Materials Science and Technology Division, Oak Ridge National Laboratory, Oak Ridge, TN, 37831, U.S.A.

Email: myoon@ornl.gov; zhangpe@msu.edu
\# These authors contributed equally.

**Abstract**

In this work, we demonstrate the formation and electronic influence of lateral heterointerfaces in FeSn containing Kagome and honeycomb layers. Lateral heterostructures offer spatially resolved property control, enabling the integration of dissimilar materials and promoting phenomena not typically observed in vertical heterostructures. Using the molecular beam epitaxy technique, we achieve a controllable synthesis of lateral heterostructures in the Kagome metal FeSn. With scanning tunneling microscopy/spectroscopy in conjunction with first-principles calculations, we provide a comprehensive understanding of the bonding motif connecting the $Fe_3Sn$-terminated Kagome and $Sn_2$-terminated honeycomb surfaces. More importantly, we reveal a distance-dependent evolution of the electronic states in the vicinity of the heterointerfaces. This evolution is significantly influenced by the orbital character of the flat bands. Our findings suggest an approach to modulate the electronic properties of the Kagome lattice, which should be beneficial for the development of future quantum devices.

An in-depth understanding of exotic quantum phases in condensed matter systems is highly imperative for the development and designs of future quantum devices relying on strong electronic correlations. Among various condensed matter systems, Kagome lattices have been quickly emerging as one of the most important platforms for studying correlated and topological electronic states.[1] In the Kagome lattice, atoms are arranged into a two-dimensional network comprising hexagons interspersed with triangles.[2] Linearly dispersing Dirac band and non-dispersing (flat) band are expected with the latter arising from the destructive interference of Bloch wave functions that leads to the self-localization of electrons in the central honeycomb ring when only nearest-neighbour hopping is considered.[3-4] Till date, the Kagome lattice has been realized in a wide range of materials, including layered materials,[4-10] non-van der Waals materials,[3, 11] and organometallic frameworks,[12-18] which are expected to lead to the realization of correlated topological phases,[10, 19-20], excitonic Bose-Einstein condensation,[21-22] spin liquid,[23-24] unconventional charge density waves,[5, 25-28] magnetically intertwined superconductivity,[1, 29-30] and spintronic devices.[31]

Among the various types of Kagome lattice materials, the Kagome metal FeSn has attracted tremendous attention. Compared to its binary $T_mX_n$ sibling compounds (e.g., $Fe_3Sn_2$ and $Fe_3Sn$), the Kagome planes in FeSn are spatially isolated from each other by the intercalation of $Sn_2$ layers, making it the compound closest to the two-dimensional limit.[32-35] Recent studies have revealed the presence of Dirac fermions and flat bands in bulk FeSn crystals using angle-resolved photoemission spectroscopy and de Haas-van Alphen quantum oscillations.[35] Theoretical studies have uncovered intricate details on the behavior of flat bands in Kagome metals.[36] In particular, it has been shown that topological flat bands arise predominantly from the $d_{z^2}$ orbital, which is consistent with the symmetries of the Kagome lattice, including its three-fold rotational symmetry. Strong crystal field splitting is essential to clearly distinguish these bands; in its absence, inter-orbital interactions can disrupt the stability of the flat band formation. Such findings contradict the widespread belief in the ubiquity of ideal flat bands in Kagome metals, since multiple $d$-orbitals with incompatible symmetries and/or insufficient crystal field splitting could prevent their manifestation.[36] Nevertheless, the potential for strong electron correlation effects remains due to the high localization of the $d$-bands. Complementary to the bulk studies, epitaxial FeSn films grown on a substrate, $SrTiO_3$ (STO) (111), were shown to exhibit surface flat bands that can potentially be integrated into heterostructures for various device applications.[37] Previous investigations have primarily focused on vertical heterointerfaces along the stacking direction. A critical gap remains in the control of the lateral heterointerfaces formed between the $Fe_3Sn$ Kagome and $Sn_2$ honeycomb layers and the thorough understanding of their influences on the electronic structures of the Kagome lattice. The study of the lateral heterointerfaces has proven challenging due to the absence of exposed $Fe_3Sn$- and $Sn_2$-terminated surfaces on the same plane in FeSn single crystals, which has limited the current understanding.

In this study, we report the formation and profound electronic implications of lateral heterostructures of $Fe_3Sn/Sn_2$ in FeSn thin films. The films are grown epitaxially on STO (111) using molecular beam epitaxy and investigated *via* the combination of scanning tunneling microscopy/spectroscopy (STM/STS) and density functional theory (DFT) calculations which provide a comprehensive approach to understanding. The epitaxial growth of FeSn films on STO allows a thorough examination of the lateral $Fe_3Sn/Sn_2$ heterointerfaces, which is not possible in bulk FeSn single crystals. Our STS results reveal three distinct density of states peaks on the $Fe_3Sn$-terminated surface. Those located at about -0.2 eV and 0.1 eV can be assigned to the surface flat bands of the Kagome lattice with $3d$ orbital character, as confirmed by the DFT calculations, whereas the peak at -0.05 eV may originate from the parabolic band by gaping out of a dispersive band of FeSn. In addition, we determine the bonding motif between the $Fe_3Sn$- and $Sn_2$-terminated surfaces at the lateral heterointerfaces. More notably, we discover an unusual long-range effect of the lateral heterointerface, where the surface flat bands are suppressed near the interface but recovered at a distance that depends on the orbital character of the state, as confirmed by STS line spectroscopy and DFT calculations. Our findings elucidate the impact of lateral heterointerfaces on the electronic behavior of the Kagome lattice, potentially providing a tool for engineering the electronic properties of FeSn to facilitate the development of future electronic and spintronic devices.

**Results and discussions**

It was previously reported that the Kagome metal FeSn is a bulk antiferromagnet with Neel temperature $T_N$ of 370 K[38]. FeSn consists of an alternating stack of two-dimensional $Fe_3Sn$ Kagome layers and $Sn_2$ honeycomb layers, as illustrated in **Figure 1a**, along the c-axis. Here, Fe spin moments within each Kagome layer are ferromagnetically aligned but antiferromagnetically coupled between adjacent layers.[39-40] By tuning the growth parameters in the molecular beam epitaxy process, we are able to grow epitaxial films of FeSn on STO (111) with island sizes ranging from 20 nm to 70 nm. The growth of FeSn on STO (111) follows the Volmer-Weber mode, which is characterized by the formation of isolated three-dimensional flat-top islands, as shown in **Figure 1b** (see **Figure S1** and **S2** in SI for details). This result is in agreement with a recent study of the growth of FeSn films on STO (111) using the MBE technique.[41] It should be noted that the distance between the two adjacent Kagome layers in the FeSn single crystal is measured to be ~ 4.4 Å. However, as shown in the STM image and the corresponding line profile in **Figure 1c-d**, the distance between the Kagome topmost layer and the $Sn_2$ underneath layer is about 2.9 Å, while that between the $Sn_2$ layer and the second Kagome layer is approximately 1.5 Å. This result is consistent with the previous reports in bulk FeSn, which supports the stacking structure of the as grown thin films.[42-43]

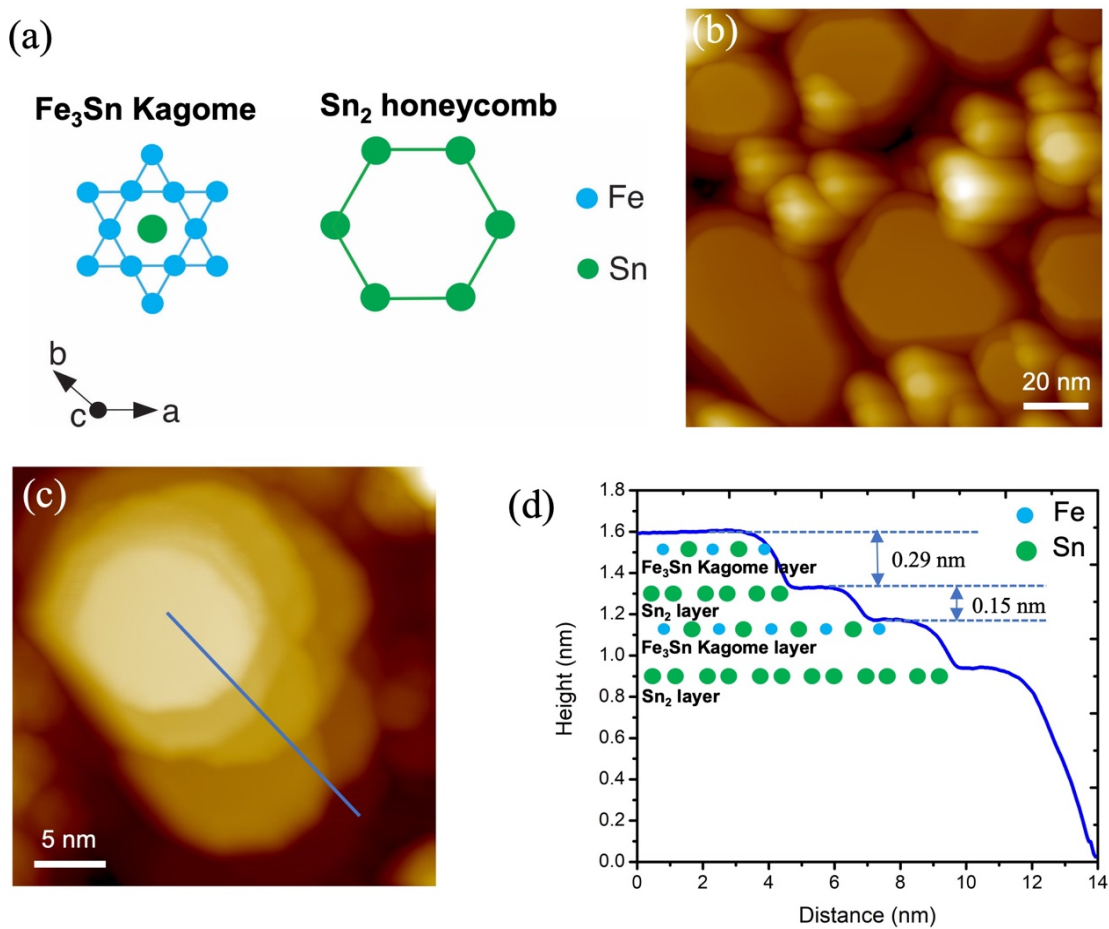

**Figure 1.** (a) In-plane schematics of the $Fe_3Sn$ Kagome- and $Sn_2$ honeycomb-terminated surfaces in FeSn. (b) Overview STM image ($V_s$ = 4 V, $I_t$ = 5 pA) showing the morphology of top-flat islands of FeSn epitaxially grown on STO (111) that follows the Volmer-Weber growth mode. (c) Close-up STM image ($V_s$ = 3 V, $I_t$ = 5 pA) illustrating the stacking of different layers in an isolated FeSn island. (d) Height profile taken along the blue line in (c), highlighting the unequal distances between the layers. The cartoon in (d) illustrates the alternating stacking of $Fe_3Sn$ Kagome and $Sn_2$ layers along the c-axis of the FeSn island.

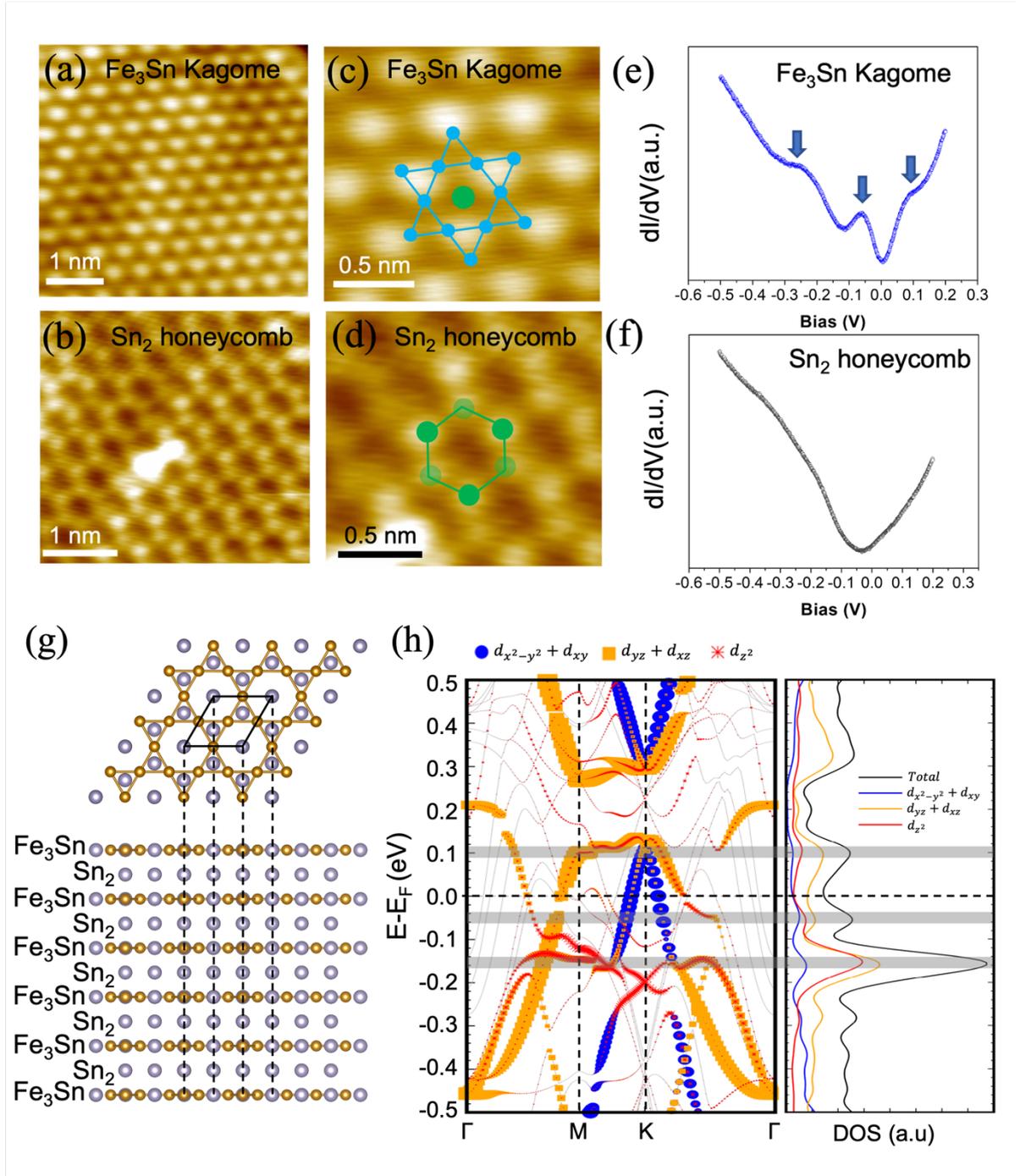

**Figure 2.** (a) A close-up STM image ($V_s$ = -0.05 V, $I_t$ = 600 pA) showing hexagonally close-packed bright protrusions of the $Fe_3Sn$-terminated Kagome surface. (b) A close-up STM image ($V_s$ = -0.05 V, $I_t$ = 500 pA) showing the buckled honeycomb pattern of the $Sn_2$-terminated surface. (c-d) Zoomed-in STM images ($V_s$ = -0.05 V, $I_t$ = 600 pA; $V_s$ = -0.05 V, $I_t$ = 500 pA) with the schematics overlaid to highlight the Star-of-David pattern of the Kagome surface and the buckled honeycomb pattern with three bright and three dark protrusions of the $Sn_2$-terminated surface, respectively. Blue dots represent Fe atom, and green dots for Sn. (e) and (f) *dI/dV* spectra (setpoint: $V_s$ = 0.3 V, $I_t$ = 260 pA) taken on the Kagome and $Sn_2$ layers, respectively. Three pronounced peaks are observed on the Kagome layer, as marked by the arrows in (e), while no significant peak is observed on the $Sn_2$ layer. (g) Top and side views of the model structure of a 6-layer (6-$Fe_3Sn$ and 6-$Sn_2$) slab, and (h) Orbital projected band structure and density of states (DOS) calculated by DFT. The highlighted gray regions represent the energies at which the peaks occur, corresponding to the arrows in (e).

**Figures 2a** and **b** show atomic resolution STM images of the Kagome- and Sn$_2$-terminated surfaces of the FeSn film grown on STO (111), respectively. The Sn$_2$-terminated surface shows a honeycomb pattern with three bright and three dark protrusions (illustrated by the green spheres of different shades in the schematics superimposed in **Figure 2d**), representing a buckled honeycomb lattice of Sn. In contrast, the Fe$_3$Sn Kagome-terminated surface exhibits hexagonally close-packed bright protrusions,[32] where the Kagome unit can be viewed as a Star-of-David pattern with a Sn atom located at the center,[32, 43] as illustrated in **Figure 2c**. To examine the electronic structures of the Kagome- and Sn$_2$-terminated surfaces, we perform the differential conductance spectra shown in **Figures 2e** and **f**, respectively. Three pronounced peaks are observed in the *dI/dV* spectrum of the Fe$_3$Sn Kagome surface: the two broad peaks are located at about -0.2 eV and 0.1 eV, while the narrow peak is located at ~ -0.05 eV. In contrast, no distinct features are found in the *dI/dV* spectrum of the Sn$_2$ layer. In addition, our *dI/dV* maps taken simultaneously with the STM images show the characteristic Kagome pattern on the Fe$_3$Sn layers in the bias range from -0.05 eV to -0.2 eV (**Figure S3** in SI).

To gain a better understanding of the origin of the peaks in the *dI/dV* spectrum of the Kagome surface, we perform DFT calculations on a 6-layer FeSn slab terminated by the Kagome layer, as shown in Figure 2g. The analysis of the orbital characters of the DOS and the band structures (see **Figure 2h**) allow us to identify the characteristics of the high intensity peaks in the experimental *dI/dV* spectrum: the surface flat band of the Kagome lattice results in a broad peak near -0.2 eV, due to the contributions from both *d$_{z2}$* and (*d$_{xz}$* + *d$_{yz}$*) orbitals of Fe atoms. The peak at 0.1 eV is attributed to the (*d$_{xz}$* + *d$_{yz}$*) orbital character, which contributes to the surface flat band of the Kagome lattice along the M-K line. It is worth noting that there are two distinct types of flat bands in the Kagome layer of FeSn: the bulk flat band and the surface flat band.[35, 37, 43-45] The bulk flat band is located at around 0.6 eV with the major contribution from the *d$_{xz}$*/*d$_{yz}$* and *d$_{xy}$*/*d$_{x2-y2}$* orbitals. Further details on the orbital characters of the electronic states of the Fe$_3$Sn Kagome layer can be found in **Figure S4**, SI. The observation that bands having significant *d$_{z2}$* contributions are wide flat band within the Brillouin zone, contrasting with those from other *d* orbitals which form narrower flat bands, aligns closely with the latest theoretical findings.[36] These findings emphasize the significance of *d*-orbital symmetry (from orbital rotation) and crystal field splitting in the formation of flat bands in Kagome lattices. As for the peak located at -0.05 eV, it could result from a gap out by the hybridization between the non-orthogonal *d$_{xz}$*/*d$_{yz}$* orbital of the Kagome surface and the bulk state (see **Figure S5** and the associated discussion in SI). Although the interlayer coupling between the consecutive Kagome planes is suppressed by the $Sn_2$ spacing layer in FeSn, hybridization between the Kagome layer and the adjacent $Sn_2$ layer can be nonnegligible and impact the electronic structures of the Kagome lattice, which is similar to the role of intercalated layers in other two-dimensional materials.[46-47]

Interestingly, we observe lateral heterointerfaces formed between the $Fe_3Sn$ Kagome and $Sn_2$-terminated surfaces (**Figure 3a**). **Figure 3b** shows a high-resolution STM image taken at the lateral heterointerface. The Kagome and honeycomb layers are identified on the lower left and upper right corners, respectively, using the characteristics of the two surfaces discussed in **Figure 2a-d**. These two layers are on the same surface plane, as confirmed by the height profile analysis (**Figure S6**, SI), which could arise from the formation of stacking faults in the epitaxial film. The arrangement of individual Fe and Sn atoms at the interface is illustrated on the STM image. The honeycomb surface displays a zigzag edge, while the Kagome surface features a straight edge. The two different lattices are covalently linked by Fe-Sn bonds in the pentagon-heptagon pairs (5-7 rings) that consist of Fe and Sn atoms, which is considered the motif driving the formation of the heterointerface.

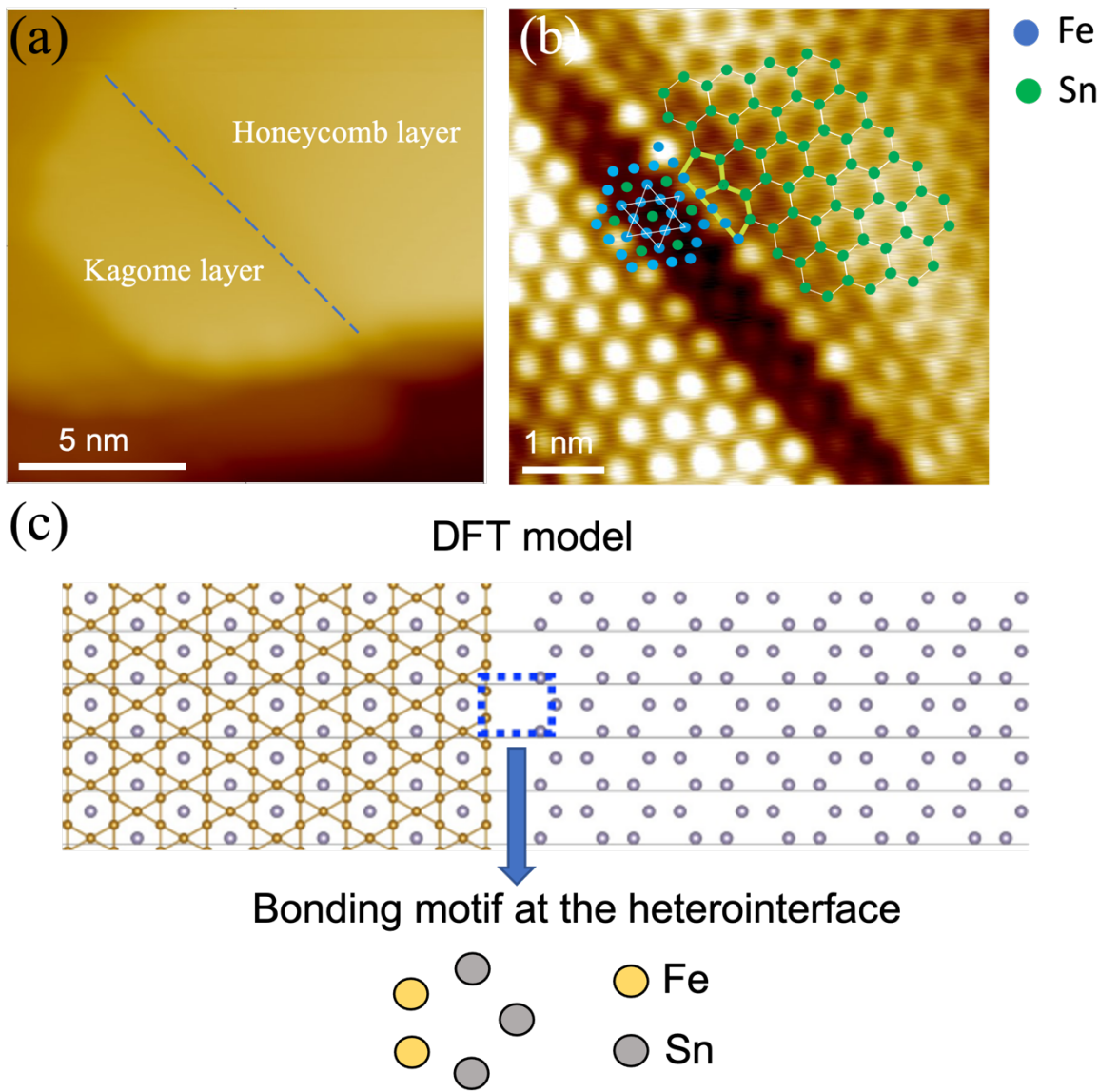

**Figure 3**. (a) STM image ($V_s$ = 3 V, $I_t$ = 5 pA) illustrating the in-plane boundary formed between the Fe$_3$Sn Kagome and Sn$_2$ honeycomb layers, as marked by the dashed blue line. (b) High-resolution STM image ($V_s$ = 3 V, $I_t$ = 5 pA) showing the atomic arrangement at the heterointerface boundary. The molecular models are overlaid on the STM image to illustrate how the Kagome and honeycomb layers are covalently linked together at the interface. (c) DFT model with the most energetically favorable bonding motif at the interface, which is consistent with that determined from the STM image in (b).

We further perform DFT calculations to gain insights into the chemical configuration of the bonding at the Kagome-honeycomb interface. **Figure 3c** shows the calculated molecular model of the boundary, which is energetically more favorable than the various other bonding possibilities, as illustrated in **Figure S7**, SI. Apparently, it fits very well with the pentagon-heptagon configuration determined from the STM images, which is responsible for the formation of the Kagome-honeycomb interface found in this work. From high-resolution STM images, we occasionally observe another type of bonding motif at the interface as depicted in **Figure S8**, SI. The presence of this configuration may be related to some intermediate states during the formation of the interface.

Furthermore, based on our line *dI/dV* spectroscopy measurements, we find that the lateral heterointerface significantly impacts the electronic structures of the Kagome lattice in areas close to the boundary. STM images in **Figure 4a** and **Figure S9b** show the FeSn island and the corresponding Kagome-honeycomb lateral heterointerface with the bonding motif identical to that previously described in **Figure 3**. Line STS spectra taken across the boundary reveal that the top right area is a $Sn_2$-terminated surface while the bottom left area is the $Fe_3Sn$ Kagome surface (see **Figure S9**, SI). **Figure 4b** depicts the trace of the STS spectra from a start point located right next to the boundary (point 1) to an endpoint located far away from the boundary on the $Fe_3Sn$ Kagome surface (point 28). Apparently, the STS curves taken right next to the boundary do not show any characteristic peaks of the $Fe_3Sn$ Kagome layer (point 1 in **Figure 4b** and the corresponding black curve in **Figure 4c**), suggesting that the heterointerface boundary suppresses the electronic structure and surface flat bands of $Fe_3Sn$. At locations far away from the boundary, the characteristic peaks of the $Fe_3Sn$ Kagome surface are recovered (point 28 in **Figure 4b** and the corresponding green curve in **Figure 4c**). Along the path, the peak located at about -0.2 eV appears first, as illustrated in the spectrum taken on point 4 (red curve in **Figure 4c**). In contrast, the peaks at ~ -0.05 eV and 0.1 eV do not emerge until further away from the boundary edge, as marked by the blue curve (point 9) in **Figure 4c**, and once established, these peaks remain nearly unchanged, i.e., from point 9 (blue dot in **Figure 4b**) to point 28 (green dot in **Figure 4b**).

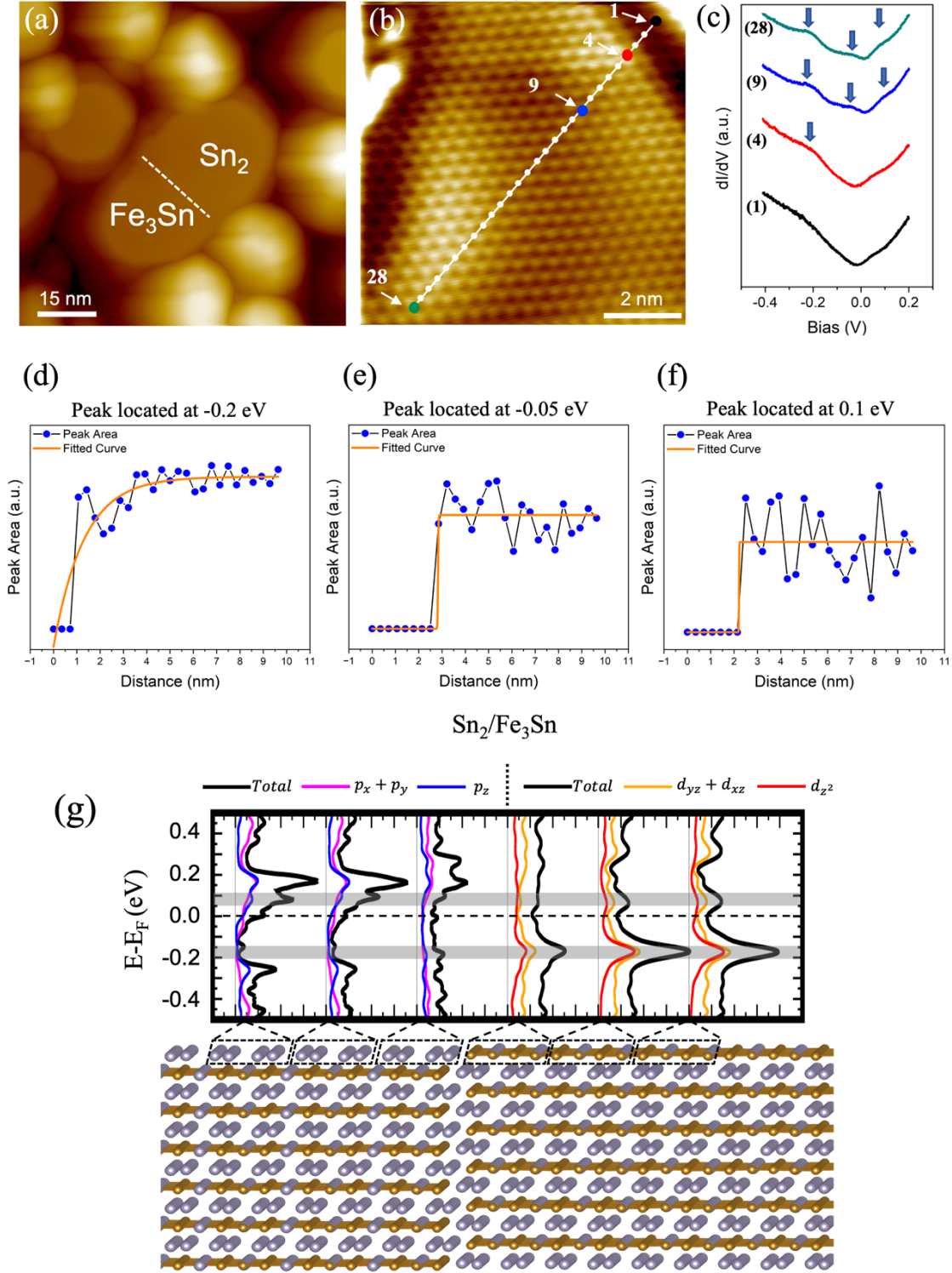

**Figure 4**. (a) Overview STM image ($V_s$ = 3 V, $I_t$ = 5 pA) illustrating the heterointerface boundary where the line STS spectra are taken. (b) Zoomed-in STM image ($V_s$ = -0.05 V, $I_t$ = 600 pA) into the Kagome surface area next to the heterointerface boundary. (c) The line STS spectra (setpoint: $V_s$ = 0.4 V, $I_t$ = 340 pA) are taken along the white trace from the heterointerface into the Kagome surface area shown in (b). (d), (e), and (f) Areas of the peaks located at ~ -0.2 eV, -0.05 eV, and 0.1 eV vs. distance from the heterointerface boundary extracted from the STS line spectra. Solid orange curves are the fittings. (g) Partial orbital projected density of states (PDOS) for the top layer atoms at the heterointerface boundary, based on the 6-layer FeSn slab structure. The highlighted gray regions represent the energies at which the peaks occur, corresponding to the arrows at -0.2 V and 0.1 V in (c).

To be quantitative on the distance-dependence of the peak evolution from the heterointerface boundary, we plot the areas of the three distinct peaks, as shown in **Figure 4d-f**, respectively, for the entire 28 STS line spectra and obtain the fitting curves. Detailed information of the analysis, as well as the fitting parameters, can be found in **Figures S10** and **S11** in SI. Our results unambiguously reveal long-range heterointerface effects on the electronic structures of the $Fe_3Sn$ lattice. The overall trends of the distance-dependent evolution are similar between the electronic states at ~ -0.05 eV and 0.1 eV, but different to that at ~ -0.2 eV, as evidenced by the different functions used in the fittings. The "coupling" length to the heterointerface is ~ 1.3 nm for the peak at -0.2 eV, while for the peaks located at -0.05 eV and 0.1 eV, this length is about 2.8 nm and 2.2 nm, respectively. Note that STM images display the convolution of geometric and electronic structures. The suppression of the electronic states near the lateral heterointerface could lead to perturbations to the morphology, e.g., the dark regions next to the boundary on the Kagome layer (**Figures 3b and 4b**).

To elucidate the long-range effects of the lateral heterointerface, we perform DFT calculations. Our model system, consisting of a lateral heterointerface connecting $Fe_3Sn$ and $Sn_2$ terminated surfaces (**Figure 3c**), is 6 layers thick (**Figure 4g**). Only a peak at ~ -0.2 eV is visible close to the lateral heterointerface, and its intensity increases as further away from the boundary. Conversely, the peak at 0.1 eV, another surface flat band feature as discussed earlier, is not visible near the boundary, appearing slightly away. This result agrees well with the distance dependence of the peak evolution observed in the experiments, which can be attributed to the orbital character of the electronic states. The peak at 0.1 eV contains the major contributions from the $(d_{xz} + d_{yz})$ orbitals. As these states with the *xy*-plane component interact with the *p* orbitals of the $Sn_2$-terminated surface at the same energy, its intensity near the lateral heterointerface disappears. In contrast, the peak at -0.2 eV is contributed by the $d_{z^2}$ and $(d_{xz} + d_{yz})$ orbitals together. The number of the *p* orbital states on the $Sn_2$-terminated surface at the same energy level is very small compared to the number of the states at 0.1 eV, therefore the interaction with the $d_{z^2}$ and $(d_{xz} + d_{yz})$ of $Fe_3Sn$ is rather weak, resulting in only a reduction instead of complete suppression of the peak at -0.2 eV near the lateral heterointerface. This observation clearly illustrates the orbital selectivity of the flat band in the lateral heterostructures. Lastly, the peak at -0.05 eV observed in the experiment is absent in the DFT results. As shown in **Figure 4d-f**, the three $Fe_3Sn$ states have different "coupling" lengths with the heterointerface. This length is longest at ~ 3 nm for the peak at -0.05 eV. Unfortunately, the lateral distance between the bulk and the heterointerface is less than 3 nm in our atomic model due to the limitation of the number of atoms in the DFT. Therefore, the peak at -0.05 eV, with the most extended interaction length with the heterointerface observed in the experiment, is not captured in DFT.

In addition, we have also identified $Fe_3Sn$-$Fe_3Sn$ and $Sn_2$-$Sn_2$ homo-interfaces formed between two different $Fe_3Sn$ or $Sn_2$ domains. Detailed information can be found in **Figures S13** and **S14** in SI. These boundaries have no pronounced influence on the electronic structures of the adjacent domains, different from the heterointerfaces. We attribute it to the formation of a 'defect'-like boundary with phase slips at the homo-interface as compared to the strong covalent bonding that leads to the influence of the $Sn_2$ electronic states on that of $Fe_3Sn$ at the 'sharp' heterointerface.

**Conclusions**

In this work, we investigate the effects of the Kagome-honeycomb lateral heterointerface on the flat bands and electronic structures of the $Fe_3Sn$ Kagome layer *via* the combination of STM/STS experiments and DFT calculations. Lateral heterointerfaces are established in the FeSn thin films epitaxially grown on the STO (111) substrate by MBE. Our STS data show the distinct peaks on the Kagome surface, whose origin and orbital characters are thoroughly explained by the DFT calculations. We further demonstrate that the Kagome-honeycomb lateral heterointerface has profound and long-range impacts on the electronic structures of the Kagome layer. Particularly, the two surface flat bands of the Kagome layer with the different orbital characters respond differently to the Kagome-honeycomb interface. This study and the orbital selectivity mechanism give rise to the potential for engineering the electronic properties of Kagome metal FeSn to facilitate the development of future quantum devices.

**Experimental method**

The substrate preparation and sample growth were performed in a standard molecular beam epitaxy (MBE) chamber with a base pressure of $6 \times 10^{-10}$ mbar. After treatment, samples were directly transferred *in-situ* into an Omicron low temperature scanning tunneling microscope (STM) operated at the liquid nitrogen temperature (~77.5K) with based pressure of $1.8 \times 10^{-11}$ mbar for characterization. Before deposition, the Nb-doped (0.5% by weight) $SrTiO_3(111)$ was cleaned using acetone and isopropyl alcohol, then immediately transferred to the MBE chamber. The substrate was slowly heated up and kept at 400 °C for 60 mins to ensure a complete degassing. The substrate was then annealed at approximately 1150 °C using direct current heating for 60 mins and then gradually cooled down to obtain a proper smooth and clean surface for the growth. FeSn was deposited onto the substrate in the MBE chamber by co-deposition of pure Sn and Fe using two different electron beam evaporators with flux current of 200 nA and 0.4 nA, respectively. The substrate was maintained at around 530 °C during the growth using resistive heating facilities. The temperature of the substrate was monitored by a thermocouple mounted at the heating stage in the MBE chamber. STS spectra were obtained using a lock-in amplifier with the modulation signal set at 26 meV in amplitude and 1.1 kHz in frequency. The STM tip was calibrated by measuring reference spectra on the silver substrate to avoid tip artifacts.

**Theoretical approaches**

We performed ab *initio* calculations based on density functional theory (DFT)[48-49] as implemented in the Vienna Ab Initio Simulation Package (VASP)[50-51] with projector augmented wave potentials[52-53] and spin polarization. The Perdew-Burke-Ernzerhof (PBE) form[54] was employed for the exchange-correlation functional with the generalized gradient approximation (GGA). The energy cutoff was set to 500 eV for all calculations. The Brillouin zone was sampled with a 25×25×1 Γ-centered k-grid. Atomic relaxations were performed until the Helmann-Feynman force acting on each atom became smaller than 0.01 eV/Å. FeSn adopts the *P6/mmm* space group, with the unit cell containing two Kagome layers composed of Fe atoms. These layers are enclosed by the Sn layer along the out-of-plane direction. The lattice parameters of bulk FeSn are found to be a = b = 5.285 Å and c = 4.444 Å. These values are in agreement with experimental measurements of a = b = 5.298 Å and c = 4.448 Å.[55]

**Acknowledgment**

Experimental work was supported by the U.S. Department of Energy (DOE), Office of Basic Energy Sciences, Division of Materials Sciences and Engineering under Award Number DE-SC0019120. P. P. Zhang acknowledges the financial support from National Science Foundation (DMR-2112691). Theory work was supported by the U.S. Department of Energy, Office of Science, Office of Basic Energy Sciences, Materials Sciences and Engineering Division (M. Yoon) and by the U.S. Department of Energy (DOE), Office of Science, National

Quantum Information Science Research Centers, Quantum Science Center (S.-H. Kang) and this research used resources of the Oak Ridge Leadership Computing Facility at the Oak Ridge National Laboratory, which is supported by the Office of Science of the U.S. Department of Energy under Contract No. DE-AC05-00OR22725 and resources of the National Energy Research Scientific Computing Center, a DOE Office of Science User Facility supported by the Office of Science of the U.S. Department of Energy under Contract No. DE-AC02-05CH11231 using NERSC award BES-ERCAP0024568.

**Supporting Information Available:** Growth morphology of FeSn thin films, STM and simultaneously obtained *dI/dV* maps obtained at different biases on the Kagome-terminated surface, Orbital origin of the Kagome layer electronic structure by DFT, Bonding motif of the Kagome-honeycomb lateral heterointerface, Quantitative analysis of the STS curves, Homo-interfaces formed between the Kagome-Kagome or honeycomb-honeycomb lattices.

**References**

1. Yin, J.-X.; Lian, B.; Hasan, M. Z., Topological Kagome Magnets and Superconductors. *Nature* **2022,** *612*, 647-657.
2. Huhtinen, K.-E.; Tylutki, M.; Kumar, P.; Vanhala, T. I.; Peotta, S.; Törmä, P., Spin-Imbalanced Pairing and Fermi Surface Deformation in Flat Bands. *Phys. Rev. B* **2018,** *97*, 214503.
3. Li, Z.; Zhuang, J.; Wang, L.; Feng, H.; Gao, Q.; Xu, X.; Hao, W.; Wang, X.; Zhang, C.; Wu, K.; Dou, S. X.; Chen, L.; Hu, Z.; Du, Y., Realization of Flat Band with Possible Nontrivial Topology in Electronic Kagome Lattice. *Sci. Adv.* **2018,** *4*, eaau4511.
4. Li, M.; Wang, Q.; Wang, G.; Yuan, Z.; Song, W.; Lou, R.; Liu, Z.; Huang, Y.; Liu, Z.; Lei, H.; Yin, Z.; Wang, S., Dirac Cone, Flat Band and Saddle Point in Kagome Magnet YMn$_6$Sn$_6$. *Nat. Commun.* **2021,** *12*, 3129.
5. Jiang, Y.-X.; Yin, J.-X.; Denner, M. M.; Shumiya, N.; Ortiz, B. R.; Xu, G.; Guguchia, Z.; He, J.; Hossain, M. S.; Liu, X.; Ruff, J.; Kautzsch, L.; Zhang, S. S.; Chang, G.; Belopolski,

I.; Zhang, Q.; Cochran, T. A.; Multer, D.; Litskevich, M.; Cheng, Z.-J., et al., Unconventional Chiral Charge Order in Kagome Superconductor KV$_3$Sb$_5$. *Nat. Mater.* **2021,** *20*, 1353-1357.

6. Zhang, W.; Liu, X.; Wang, L.; Tsang, C. W.; Wang, Z.; Lam, S. T.; Wang, W.; Xie, J.; Zhou, X.; Zhao, Y.; Wang, S.; Tallon, J.; Lai, K. T.; Goh, S. K., Nodeless Superconductivity in Kagome Metal CsV$_3$Sb$_5$ with and without Time Reversal Symmetry Breaking. *Nano Lett.* **2023,** *23*, 872-879.

7. Ortiz, B. R.; Gomes, L. C.; Morey, J. R.; Winiarski, M.; Bordelon, M.; Mangum, J. S.; Oswald, I. W. H.; Rodriguez-Rivera, J. A.; Neilson, J. R.; Wilson, S. D.; Ertekin, E.; McQueen, T. M.; Toberer, E. S., New Kagome Prototype Materials: Discovery of KV$_3$Sb$_5$, RbV$_3$Sb$_5$, and CsV$_3$Sb$_5$. *Phys. Rev. Mater.* **2019,** *3*, 094407.

8. Morali, N.; Batabyal, R.; Nag, P. K.; Liu, E.; Xu, Q.; Sun, Y.; Yan, B.; Felser, C.; Avraham, N.; Beidenkopf, H., Fermi-Arc Diversity on Surface Terminations of the Magnetic Weyl Semimetal Co$_3$Sn$_2$S$_2$. *Science* **2019,** *365*, 1286-1291.

9. Dally, R. L.; Lynn, J. W.; Ghimire, N. J.; Michel, D.; Siegfried, P.; Mazin, I. I., Chiral Properties of the Zero-Field Spiral State and Field-Induced Magnetic Phases of the Itinerant Kagome Metal YMn$_6$Sn$_6$. *Phys. Rev. B* **2021,** *103*, 094413.

10. Yin, J.-X.; Ma, W.; Cochran, T. A.; Xu, X.; Zhang, S. S.; Tien, H.-J.; Shumiya, N.; Cheng, G.; Jiang, K.; Lian, B.; Song, Z.; Chang, G.; Belopolski, I.; Multer, D.; Litskevich, M.; Cheng, Z.-J.; Yang, X. P.; Swidler, B.; Zhou, H.; Lin, H., et al., Quantum-Limit Chern Topological Magnetism in TbMn$_6$Sn$_6$. *Nature* **2020,** *583*, 533-536.

11. Han, T.-H.; Helton, J. S.; Chu, S.; Nocera, D. G.; Rodriguez-Rivera, J. A.; Broholm, C.; Lee, Y. S., Fractionalized Excitations in the Spin-Liquid State of a Kagome-Lattice Antiferromagnet. *Nature* **2012,** *492*, 406-410.

12. Takenaka, T.; Ishihara, K.; Roppongi, M.; Miao, Y.; Mizukami, Y.; Makita, T.; Tsurumi, J.; Watanabe, S.; Takeya, J.; Yamashita, M.; Torizuka, K.; Uwatoko, Y.; Sasaki, T.; Huang, X.; Xu, W.; Zhu, D.; Su, N.; Cheng, J. G.; Shibauchi, T.; Hashimoto, K., Strongly Correlated Superconductivity in a Copper-Based Metal-Organic Framework with a Perfect Kagome Lattice. *Sci. Adv.* **2021,** *7*, eabf3996.

13. Yamada, M. G.; Soejima, T.; Tsuji, N.; Hirai, D.; Dincă, M.; Aoki, H., First-Principles Design of a Half-Filled Flat Band of the Kagome Lattice in Two-Dimensional Metal-Organic Frameworks. *Phys. Rev. B* **2016,** *94*, 081102.

14. Galeotti, G.; De Marchi, F.; Hamzehpoor, E.; MacLean, O.; Rajeswara Rao, M.; Chen, Y.; Besteiro, L. V.; Dettmann, D.; Ferrari, L.; Frezza, F.; Sheverdyaeva, P. M.; Liu, R.; Kundu, A. K.; Moras, P.; Ebrahimi, M.; Gallagher, M. C.; Rosei, F.; Perepichka, D. F.; Contini, G., Synthesis of Mesoscale Ordered Two-Dimensional π-Conjugated Polymers with Semiconducting Properties. *Nat. Mater.* **2020,** *19*, 874-880.

15. Wang, Z. F.; Su, N.; Liu, F., Prediction of a Two-Dimensional Organic Topological Insulator. *Nano Lett.* **2013,** *13*, 2842-2845.

16. Kambe, T.; Sakamoto, R.; Hoshiko, K.; Takada, K.; Miyachi, M.; Ryu, J.-H.; Sasaki, S.; Kim, J.; Nakazato, K.; Takata, M.; Nishihara, H., π-Conjugated Nickel Bis(dithiolene) Complex Nanosheet. *J. Am. Chem. Soc.* **2013,** *135*, 2462-2465.

17. Pan, M.; Zhang, X.; Zhou, Y.; Wang, P.; Bian, Q.; Liu, H.; Wang, X.; Li, X.; Chen, A.; Lei, X.; Li, S.; Cheng, Z.; Shao, Z.; Ding, H.; Gao, J.; Li, F.; Liu, F., Growth of Mesoscale Ordered Two-Dimensional Hydrogen-Bond Organic Framework with the Observation of Flat Band. *Phys. Rev. Lett.* **2023,** *130*, 036203.

18. Jiang, W.; Ni, X.; Liu, F., Exotic Topological Bands and Quantum States in Metal–Organic and Covalent–Organic Frameworks. *Acc. Chem. Res.* **2021,** *54*, 416-426.

19. Shi, M.; Yu, F.; Yang, Y.; Meng, F.; Lei, B.; Luo, Y.; Sun, Z.; He, J.; Wang, R.; Jiang, Z.; Liu, Z.; Shen, D.; Wu, T.; Wang, Z.; Xiang, Z.; Ying, J.; Chen, X., A New Class of Bilayer

Kagome Lattice Compounds with Dirac Nodal Lines and Pressure-Induced Superconductivity. *Nat. Commun.* **2022,** *13*, 2773.
20. Ye, L.; Kang, M.; Liu, J.; von Cube, F.; Wicker, C. R.; Suzuki, T.; Jozwiak, C.; Bostwick, A.; Rotenberg, E.; Bell, D. C.; Fu, L.; Comin, R.; Checkelsky, J. G., Massive Dirac Fermions in a Ferromagnetic Kagome Metal. *Nature* **2018,** *555*, 638-642.
21. Sethi, G.; Zhou, Y.; Zhu, L.; Yang, L.; Liu, F., Flat-Band-Enabled Triplet Excitonic Insulator in a Diatomic Kagome Lattice. *Phys. Rev. Lett.* **2021,** *126*, 196403.
22. Sethi, G.; Cuma, M.; Liu, F., Excitonic Condensate in Flat Valence and Conduction Bands of Opposite Chirality. *Phys. Rev. Lett.* **2023,** *130*, 186401.
23. Ko, W.-H.; Lee, P. A.; Wen, X.-G., Doped Kagome System as Exotic Superconductor. *Phys. Rev. B* **2009,** *79*, 214502.
24. Fujihala, M.; Morita, K.; Mole, R.; Mitsuda, S.; Tohyama, T.; Yano, S.-i.; Yu, D.; Sota, S.; Kuwai, T.; Koda, A.; Okabe, H.; Lee, H.; Itoh, S.; Hawai, T.; Masuda, T.; Sagayama, H.; Matsuo, A.; Kindo, K.; Ohira-Kawamura, S.; Nakajima, K., Gapless Spin Liquid in a Square-Kagome Lattice Antiferromagnet. *Nat. Commun.* **2020,** *11*, 3429.
25. Teng, X.; Chen, L.; Ye, F.; Rosenberg, E.; Liu, Z.; Yin, J.-X.; Jiang, Y.-X.; Oh, J. S.; Hasan, M. Z.; Neubauer, K. J.; Gao, B.; Xie, Y.; Hashimoto, M.; Lu, D.; Jozwiak, C.; Bostwick, A.; Rotenberg, E.; Birgeneau, R. J.; Chu, J.-H.; Yi, M., et al., Discovery of Charge Density Wave in a Kagome Lattice Antiferromagnet. *Nature* **2022,** *609*, 490-495.
26. Arachchige, H. W. S.; Meier, W. R.; Marshall, M.; Matsuoka, T.; Xue, R.; McGuire, M. A.; Hermann, R. P.; Cao, H.; Mandrus, D., Charge Density Wave in Kagome Lattice Intermetallic $ScV_6Sn_6$. *Phys. Rev. Lett.* **2022,** *129*, 216402.
27. Kang, S. H.; Li, H.; Meier, W. R.; Villanova, J. W.; Hus, S.; Jeon, H.; Arachchige, H. W. S.; Lu, Q.; Gai, Z.; Denlinger, J.; Moore, R.; Yoon, M.; Mandrus, D., Emergence of a New Band and the Lifshitz Transition in Kagome Metal $ScV_6Sn_6$ with Charge Density Wave. **2023,** *arXiv: 2302.14041*. DOI: 10.48550/arXiv.2302.14041 (accessed February 27, 2023).
28. Mozaffari, S.; Meier, W. R.; Madhogaria, R. P.; Peshcherenko, N.; Kang, S. H.; Villanova, J. W.; Arachchige, H. W. S.; Zheng, G.; Zhu, Y.; Chen, K. W.; Jenkins, K.; Zhang, D.; Chan, A.; Li, L.; Yoon, M.; Zhang, Y.; Mandrus, D., Universal Sublinear Resistivity in Vanadium Kagome Materials Hosting Charge Density Waves. **2023,** *arXiv: 2305.02393*. DOI: 10.48550/arXiv.2305.02393 (accessed May 3, 2023).
29. Chen, H.; Yang, H.; Hu, B.; Zhao, Z.; Yuan, J.; Xing, Y.; Qian, G.; Huang, Z.; Li, G.; Ye, Y.; Ma, S.; Ni, S.; Zhang, H.; Yin, Q.; Gong, C.; Tu, Z.; Lei, H.; Tan, H.; Zhou, S.; Shen, C., et al., Roton Pair Density Wave in a Strong-Coupling Kagome Superconductor. *Nature* **2021,** *599*, 222-228.
30. Neupert, T.; Denner, M. M.; Yin, J.-X.; Thomale, R.; Hasan, M. Z., Charge Order and Superconductivity in Kagome Materials. *Nat. Phys.* **2022,** *18*, 137-143.
31. Tsai, H.; Higo, T.; Kondou, K.; Nomoto, T.; Sakai, A.; Kobayashi, A.; Nakano, T.; Yakushiji, K.; Arita, R.; Miwa, S.; Otani, Y.; Nakatsuji, S., Electrical Manipulation of a Topological Antiferromagnetic State. *Nature* **2020,** *580*, 608-613.
32. Yin, J.-X.; Zhang, S. S.; Li, H.; Jiang, K.; Chang, G.; Zhang, B.; Lian, B.; Xiang, C.; Belopolski, I.; Zheng, H.; Cochran, T. A.; Xu, S.-Y.; Bian, G.; Liu, K.; Chang, T.-R.; Lin, H.; Lu, Z.-Y.; Wang, Z.; Jia, S.; Wang, W., et al., Giant and Anisotropic Many-Body Spin–Orbit Tunability in a Strongly Correlated Kagome Magnet. *Nature* **2018,** *562*, 91-95.
33. Zhang, H.; Feng, H.; Xu, X.; Hao, W.; Du, Y., Recent Progress on 2D Kagome Magnets: Binary $TmSn_n$ (T = Fe, Co, Mn). *Adv. Quantum Technol.* **2021,** *4*, 2100073.
34. Sales, B. C.; Yan, J.; Meier, W. R.; Christianson, A. D.; Okamoto, S.; McGuire, M. A., Electronic, Magnetic, and Thermodynamic Properties of the Kagome Layer Compound FeSn. *Phys. Rev. Mater.* **2019,** *3*, 114203.

35. Kang, M.; Ye, L.; Fang, S.; You, J.-S.; Levitan, A.; Han, M.; Facio, J. I.; Jozwiak, C.; Bostwick, A.; Rotenberg, E.; Chan, M. K.; McDonald, R. D.; Graf, D.; Kaznatcheev, K.; Vescovo, E.; Bell, D. C.; Kaxiras, E.; van den Brink, J.; Richter, M.; Prasad Ghimire, M., et al., Dirac Fermions and Flat Bands in the Ideal Kagome Metal FeSn. *Nat. Mater.* **2020**, *19*, 163-169.
36. Kim, D.; Liu, F., Realization of Flat Bands by Lattice Intercalation in Kagome Metals. *Phys. Rev. B* **2023**, *107*, 205130.
37. Han, M.; Inoue, H.; Fang, S.; John, C.; Ye, L.; Chan, M. K.; Graf, D.; Suzuki, T.; Ghimire, M. P.; Cho, W. J.; Kaxiras, E.; Checkelsky, J. G., Evidence of Two-Dimensional Flat Band at the Surface of Antiferromagnetic Kagome Metal FeSn. *Nat. Commun.* **2021**, *12*, 5345.
38. Yamamoto, H., Mössbauer Effect Measurement of Intermetallic Compounds in Iron-Tin System : $Fe_5Sn_3$ and FeSn. *J. Phys. Soc. Jpn.* **1966,** *21*, 1058-1062.
39. Hartmann, O.; Wäppling, R., Muon Spin Precession in the Hexagonal Antiferromagnet FeSn. *Phys. Scr.* **1987,** *35*, 499.
40. Häggström, L.; Ericsson, T.; Wäppling, R.; Chandra, K., Studies of the Magnetic Structure of FeSn Using the Mössbauer Effect. *Phys. Scr.* **1975,** *11*, 47.
41. Zhang, H.; Weinert, M.; Li, L., Giant Periodic Pseudomagnetic Fields in Strained Kagome Magnet FeSn Epitaxial Films on $SrTiO_3$(111) Substrate. *Nano Lett.* **2023,** *23*, 2397-2404.
42. Li, H.; Zhao, H.; Yin, Q.; Wang, Q.; Ren, Z.; Sharma, S.; Lei, H.; Wang, Z.; Zeljkovic, I., Spin-Polarized Imaging of the Antiferromagnetic Structure and Field-Tunable Bound States in Kagome Magnet FeSn. *Sci. Rep.* **2022,** *12*, 14525.
43. Lee, S.-H.; Kim, Y.; Cho, B.; Park, J.; Kim, M.-S.; Park, K.; Jeon, H.; Jung, M.; Park, K.; Lee, J.; Seo, J., Spin-Polarized and Possible Pseudospin-Polarized Scanning Tunneling Microscopy in Kagome Metal FeSn. *Commun. Phys.* **2022,** *5*, 235.
44. Multer, D.; Yin, J.-X.; Hossain, M. S.; Yang, X.; Sales, B. C.; Miao, H.; Meier, W. R.; Jiang, Y.-X.; Xie, Y.; Dai, P.; Liu, J.; Deng, H.; Lei, H.; Lian, B.; Zahid Hasan, M., Imaging Real-Space Flat Band Localization in Kagome Magnet FeSn. *Commun. Mater.* **2023,** *4*, 17.
45. Kang, M.; Fang, S.; Ye, L.; Po, H. C.; Denlinger, J.; Jozwiak, C.; Bostwick, A.; Rotenberg, E.; Kaxiras, E.; Checkelsky, J. G.; Comin, R., Topological Flat Bands in Frustrated Kagome Lattice CoSn. *Nat. Commun.* **2020,** *11*, 4004.
46. Chen, S.; Han, Y.; Kolmer, M.; Hall, J.; Hupalo, M.; Evans, J. W.; Tringides, M. C., Targeted Dy Intercalation under Graphene/SiC for Tuning its Electronic Band Structure. *Phys. Rev. B* **2023,** *107*, 045408.
47. Link, S.; Forti, S.; Stöhr, A.; Küster, K.; Rösner, M.; Hirschmeier, D.; Chen, C.; Avila, J.; Asensio, M. C.; Zakharov, A. A.; Wehling, T. O.; Lichtenstein, A. I.; Katsnelson, M. I.; Starke, U., Introducing Strong Correlation Effects into Graphene by Gadolinium Intercalation. *Phys. Rev. B* **2019,** *100*, 121407.
48. Kohn, W.; Sham, L. J., Self-Consistent Equations Including Exchange and Correlation Effects. *Phys. Rev.* **1965,** *140*, A1133-A1138.
49. Hohenberg, P.; Kohn, W., Inhomogeneous Electron Gas. *Phys. Rev.* **1964,** *136*, B864-B871.
50. Kresse, G.; Hafner, J., Ab initio Molecular Dynamics for Liquid Metals. *Phys. Rev. B* **1993,** *47*, 558-561.
51. Kresse, G.; Furthmüller, J., Efficient Iterative Schemes for ab initio Total-Energy Calculations Using a Plane-Wave Basis Set. *Phys. Rev. B* **1996,** *54*, 11169-11186.
52. Blöchl, P. E., Projector Augmented-Wave Method. *Phys. Rev. B* **1994,** *50*, 17953-17979.
53. Kresse, G.; Joubert, D., From Ultrasoft Pseudopotentials to the Projector Augmented-Wave Method. *Phys. Rev. B* **1999,** *59*, 1758-1775.

54. Perdew, J. P.; Burke, K.; Ernzerhof, M., Generalized Gradient Approximation Made Simple. *Phys. Rev. Lett.* **1996,** *77*, 3865-3868.
55. Yamaguchi, K.; Watanabe, H., Neutron Diffraction Study of FeSn. *J. Phys. Soc. Jpn.* **1967,** *22*, 1210-1213.